\documentstyle[twoside,fleqn,espcrc2,epsf]{article}

\newcommand{\mystrut}{\vrule height 12pt depth 5pt width 0pt}

\newcommand{\AmS}{{\protect\the\textfont2
  A\kern-.1667em\lower.5ex\hbox{M}\kern-.125emS}}

\def\etal{{\it et al}.,\ }
\title{Light hadron spectrum---MILC results with the Kogut-Susskind and
Wilson actions\footnotemark}

\author{ Claude~Bernard,\hskip-0.03in
\address{Department of Physics, Washington University, St.~Louis, MO 63130, USA} 
Tom~Blum,\hskip-0.03in
\address{Department of Physics, Brookhaven National Lab, Upton, NY 11973, USA} 
Carleton~DeTar,\hskip-0.03in
\address{Physics Department, University of Utah, Salt Lake City, UT 84112, USA}
Steven~Gottlieb,\hskip-0.03in
\address{Department of Physics, Indiana University, Bloomington, IN 47405, USA}
Urs~M.~Heller,\hskip-0.03in
\address{SCRI, Florida State University, Tallahassee, FL 32306-4052, USA} 
Jim~Hetrick,\hskip-0.03in$\,\null^{\rm a}$
Craig~McNeile,\hskip-0.03in$\,\null^{\rm c}$
Kari~Rummukainen,\hskip-0.03in
\address{Universit\"at Bielefeld, Fakult\"at f\"ur Physik, Postfach
100131, D-33501 Bielefeld, Germany} 
Bob~Sugar,\hskip-0.03in
\address{Department of Physics, University of California, Santa Barbara, CA 93106, USA} 
Doug~Toussaint,\hskip-0.03in
\address{Department of Physics, University of Arizona, Tucson, AZ 85721, USA} 
and Matthew~Wingate
\address{Physics Department, University of Colorado, Boulder, CO 80309, USA} 
} 

\begin{document}

\begin{abstract}
We present the current status of our ongoing calculations of the light
hadron spectrum with both Kogut-Susskind (KS) and Wilson quarks in the 
valence or quenched approximation.  We discuss KS quarks first and find
that the chiral extrapolation is potentially the biggest source of
systematic error.  For the Wilson case, we focus on finite volume and
source size effects at $6/g^2=5.7$.  We find no evidence to support the
claim that there is a finite volume effect between $N_s=16$ and 24 of 
approximately 5\%.  

\end{abstract}

\maketitle

\section{INTRODUCTION}
\renewcommand{\thefootnote}{\fnsymbol{footnote}}
\footnotetext{Invited talk presented by S.~Gottlieb at ``Lattice QCD on 
Parallel Computers,'' University of Tsukuba, March, 1997, to appear in the
proceedings.}

It has been some 15 years since the first calculations of the light quark
spectrum were done on computers capable of sustaining about 1 Megaflop
\cite{PIONEERS}.  
Today, computers exist that can achieve over one million times that speed,
and if we can just get enough time on them, we should be able to control 
carefully the systematic errors arising from finite volume, a quark mass
that is heavier than in nature, and a nonzero lattice spacing.  
We are certainly close to that goal within the valence or
quenched approximation (even without going to improved lattice
actions), if not yet with dynamical
quarks.  Here we will concentrate on some recent results of the MILC
Collaboration within the quenched approximation.  We are also doing
calculations with dynamical KS quarks that are covered in another talk at
this conference \cite{TSUKUBADYN}.  
One of us has recently reviewed the light quark spectrum
at Lattice 96 \cite{SGSPECREVIEW} 
and an overview of other groups' work can be found in the
proceedings.  Additional details are available at the URL
{\tt http://physics.indiana.edu/\~{}sg/lat96\_spec trum.html}.

In the rest of this paper, 
we list the lattices that were used in our KS calculations and then discuss
the sources of systematic errors.  We first discuss what volume is needed
to avoid finite size effects, then the quark mass (or
chiral) extrapolation required
and, finally, the lattice spacing extrapolation.  We
have found that the chiral extrapolation is the most
delicate issue.  We then turn to recent calculations with Wilson
quarks.  We discuss the same sources of systematic errors and how
our calculations can help shed light on the crucial issues.
At the time of the conference, our running was not completed.  We
present some preliminary results and outline what we will be able to study
when all of our hadron propagators are analyzed.

\begin{table}[tb]
\setlength{\tabcolsep}{1.5pc}
\newlength{\digitwidth} \settowidth{\digitwidth}{\rm 0}
\catcode`?=\active \def?{\kern\digitwidth}
\caption{Lattices used for quenched Kogut-Susskind calculations}
\label{tab:kslattices}
\begin{tabular}{lccc}
\hline
\mystrut $6/g^2$&size&\#\\
\hline
\mystrut $ 5.54 $&   $16^3 \times 32$&          200 \cr
\noalign{\smallskip}
 $ 5.7 $&   $8^3 \times 48$&          600 \cr
 $ 5.7 $&   $12^3 \times 48$&          400 \cr
 $ 5.7 $&   $16^3 \times 48$&          400 \cr
 $ 5.7 $&   $20^3 \times 48$&          200 \cr
 $ 5.7 $&   $24^3 \times 48$&          200 \cr
\noalign{\smallskip}
 $ 5.85 $&   $12^3 \times 48$&          200 \cr
 $ 5.85 $&   $20^3 \times 48$&          200 \cr
 $ 5.85 $&   $24^3 \times 48$&          200 \cr
\noalign{\smallskip}
 $ 6.15 $&   $32^3 \times 64$&          135 \cr
\hline
\end{tabular}
\end{table}

\section{THE QUENCHED KS SPECTRUM}

\subsection{Ensemble and calculational details}
Table \ref{tab:kslattices} lists the couplings, volumes and number of
lattices used in our Kogut-Susskind calculation.
To generate the gauge configurations, we use a combination of the overrelaxed
algorithm \cite{OVERREL}
and the pseudo heat bath algorithm \cite{HEATBATH}
in the ratio of
4 sweeps to 1.  Each lattice is separated from the previous one
by 200 such updates, or a total of 800 overrelaxed and 200 heat bath
sweeps.  We have looked at five quark masses for each
gauge coupling.  The heaviest quark mass is 16 times the lightest.
For $6/g^2=5.54$, the range of quark mass $am_q$ is 0.02--0.32.
At 5.7, it is 0.01--0.16.  The same range is used at 5.85, and the range is
reduced to 0.005--0.08 at 6.15.

To calculate the hadron propagators, we use wall sources.  
On the source time slice, all sites with all spatial
coordinates even are set to one, all other sites to zero.
On each lattice, we have a source every 8 time slices, {\it i.e.},  
we have
four sets of propagators per lattice for $6/g^2=5.54$, six for 5.7 and
5.85 and 8 for 6.15.  We have over 1000 propagators for every case
except for 5.54.  Propagators on each lattice with different
source times are blocked together for further analysis.

\begin{figure}[thb]
\epsfxsize=0.99 \hsize
\epsffile{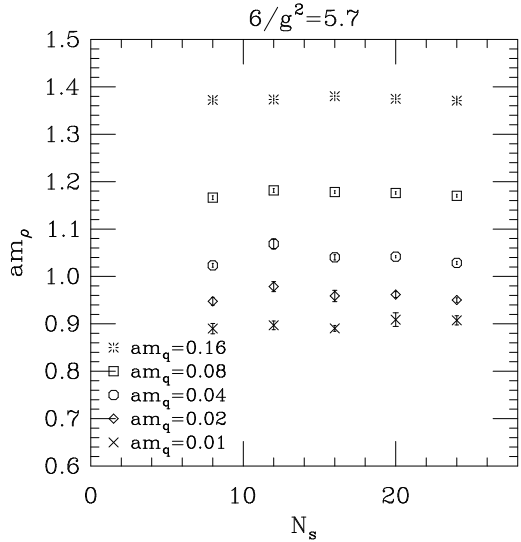}
\vspace{-28pt}
\caption{The rho mass as a function of lattice size $N_s$ for 
various quark masses with $6/g^2=5.7$.}
\label{fig:rho57}
\end{figure}

\begin{figure}[thb]
\epsfxsize=0.99 \hsize 
\epsffile{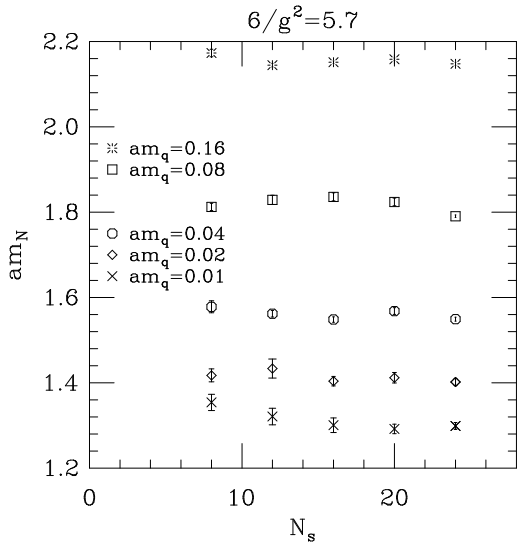} 
\vspace{-28pt}  
\caption{The nucleon mass as a function of lattice size $N_s$ for 
various quark masses with $6/g^2=5.7$.}
\label{fig:nuc57}
\end{figure}

\subsection{Finite size effects}
It is well known that if the lattice size is too small hadrons get squeezed
and their masses increase \cite{FINITESIZE}.  
For coupling 5.7 we have five lattice
sizes 8, 12, 16, 20 and 24.  In Figs.~\ref{fig:rho57} and
\ref{fig:nuc57}, we display our result for the nucleon and rho masses as a
function of spatial size in lattice units.  Over this range of sizes we do
not see any noticeable trend for the rho masses.
For the smaller quark masses, the nucleon mass
appears to be decreasing with increasing
size at $N_s=8$ and perhaps also at 12.
If we use the rho mass (extrapolated to zero quark mass)
to set the scale, the lattice spacing $a$
is 0.23 fm and this range of lattice sizes corresponds to 1.84--5.52 fm.

\begin{figure}[thb]
\epsfxsize=0.99 \hsize
\epsffile{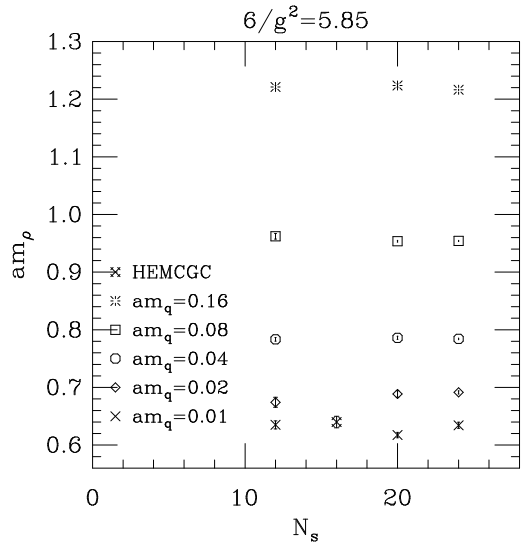}
\vspace{-28pt}
\caption{The rho mass as a function of lattice size $N_s$ for 
various quark masses with $6/g^2=5.85$.}
\label{fig:rho585}
\end{figure}

\begin{figure}[thb]
\epsfxsize=0.99 \hsize 
\epsffile{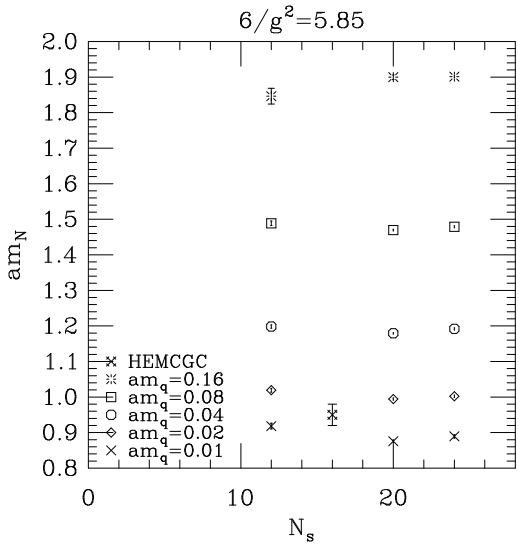} 
\vspace{-28pt}  
\caption{The nucleon mass as a function of lattice size $N_s$ for 
various quark masses with $6/g^2=5.85$.}
\label{fig:nuc585}
\end{figure}

At 5.85, we have studied three spatial sizes $N_s=12$, 20 and 24.  
(See Figs.~\ref{fig:rho585} and \ref{fig:nuc585}.)  Again, we see
a similar picture with the nucleon showing a discernible effect for the
lightest two quark masses.  With a lattice spacing of about 0.15 fm, these sizes
correspond to 1.8, 3.0 and 3.6 fm.  With some evidence of where finite
size effects at 5.7 and 5.85 become small,
we can now appeal to the fact that this
effect is physical.  We argue that at 6.15 we have a large enough box size
that we can safely ignore finite volume corrections.
The lattice spacing at 6.15 is about 0.085 fm, so our box size is about
2.7 fm.


\subsection{Chiral extrapolations}

We have found that the chiral extrapolation is the most delicate issue in
controlling the possible systematic errors.  
The extrapolation is complicated by several factors.  First, the chiral
expansion is a small mass expansion, but the most precise results we can
generate are for large quark mass.  As we reduce
the quark mass, the statistical
errors in the hadron masses grow.  We would like to start
our chiral fits in the small mass region where they are simplest and
most reliable.  We could then include results for heavier masses and see
when higher order terms in the expansion are required.  Instead,
we have to apply our fits in a region where there may be large
contributions from higher order terms.  If our hadron masses are not very
precise, or our range of quark mass is not very wide, a simple linear fit
will be possible, but it may have little to do with the true low quark
mass limit.  The second complication is due to the fact that quenched
chiral perturbation theory ($Q\chi PT$) \cite{QCPT}
is different from ordinary 
chiral perturbation theory ($\chi PT$) that applies to the real world
\cite{CPT}.  The differences
between quenched and ordinary chiral perturbation theory are based on
continuum rather than lattice calculations.  The size of the additional
terms is not well determined and they have not been clearly seen in the
lattice results \cite{CHIRALRESULTS}.
The third complication involves possible systematic effects in fitting 
the hadron propagators.  The propagators must not be fit too close to the
source plane or we may not be in the asymptotic region.  However,
there is a strong inducement in terms of statistical accuracy to fit near
the source plane.  With the wall sources we use, and with Gaussian
sources of moderate size, the masses approach their asymptotic value from
below.  Lighter mass channels can be fit closer to the source plane
because the heavier states decay away relatively more quickly; however,
we must try to guard against systematic bias (from fitting too close
to the source) that would make the hadron
masses drop too quickly as the quark mass decreases.

\begin{figure}[thb]
\epsfxsize=0.99 \hsize 
\epsffile{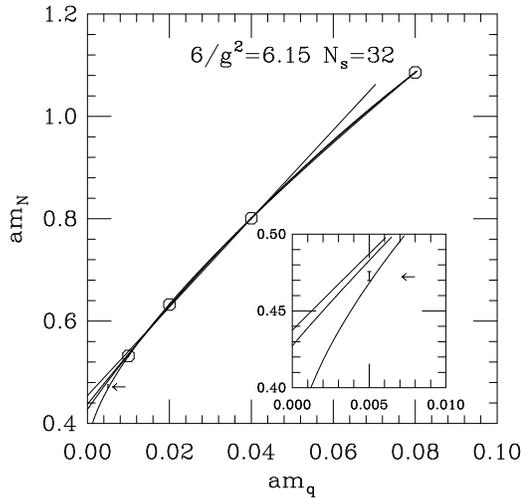} 
\vspace{-28pt}  
\caption{The chiral extrapolation for the nucleon with $6/g^2=6.15$.  
Octagons denote the points included in the fit.
An arrow points to the lightest mass result which was not part of the fit.}
\label{fig:nuc6pt15}
\end{figure}

We have five quark masses in each of our calculations.  We cannot
include too many higher order terms in the chiral expansion and still have
any degrees of freedom left in the fit.  
In Fig.~\ref{fig:nuc6pt15}, we display the nucleon mass 
and some simple fits.  There is
clear curvature in the data.  In fact, no three points can be adequately
fit by a straight line.  
Non-linear terms are required in the chiral fit, and since the error
on the lightest mass point is largest, we attempt some nonlinear, three
parameter fits to the four heaviest masses.
Each curve includes one additional power of quark mass.
Looking at small quark mass, the upper two curves include $m_q^{3/2}$ or
$m_q^2$.  These are higher order terms that occur in ordinary $\chi PT$.
The lowest curve includes the power $m_q^{1/2}$, a
term that only appears in $Q\chi PT$.
In this case, the curve with the square root very nicely touches the
error bar of
the point that was not part of the fit.  

\begin{table}
\caption{Our fitting functions}
\label{tab:fittingforms}
\setlength{\tabcolsep}{1.1pc}
\begin{tabular}{ll}
\hline
\mystrut\rm Fit 1:&$M+ a m^{1/2}$\\
\rm Fit 2:&$M+ a m^{1/2} + b m$\\
\rm Fit 3:&$M+ a m^{1/2} + b m +c m^{3/2}$\\
\rm Fit 4:&$M+ a m^{1/2} + b m +c m^2$\\
\rm Fit 5:&$M+ a m$\\
\rm Fit 6:&$M+ a m  +b m^{3/2}$\\
\rm Fit 7:&$M+ a m  +b m^2$\\
\rm Fit 8:&$M+ a m  +b m^{3/2} + c m^2$\\
\rm Fit 9:&$M+ a m  +b m \log m$\\
\rm Fit 10:&$M+ a m^{1/2} + b m +c m\log m$ \\
\rm Fit 11:&$M+ a m  +b m^2 \log m$\\
\rm Fit 12:&$M+ a m  +b m^2 + c m^2 \log m$\\
\hline
\end{tabular}
\end{table}

\begin{figure}[thb]
\epsfxsize=0.99 \hsize
\epsffile{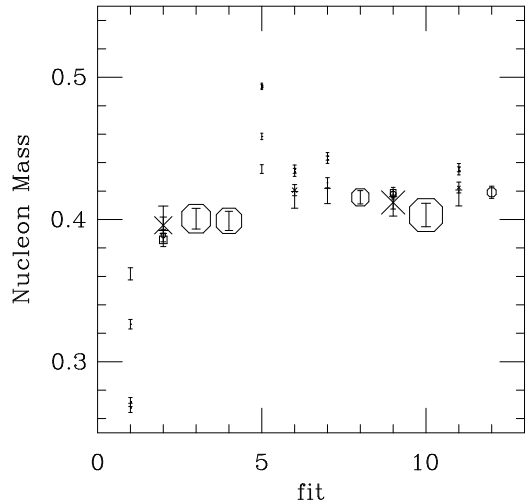}
\vspace{-28pt}
\caption{The nucleon mass for $6/g^2=6.15$
extrapolated to the physical quark mass (as determine by $m_\pi/m_\rho$)
for various
fits.  The symbol sizes are proportional to the confidence level of the fits.
The meaning of the symbols is described in the text.}
\label{fig:nucmphysall}
\end{figure}

It is perhaps fortuitous that one of our three parameter fits touches the
point we did not include in the fit.  When we try to fit our other cases,
we find that the three parameter fits do not work well in all cases.
We have considered a
dozen different fitting functions with up to four parameters.  They are
listed in Table~\ref{tab:fittingforms}.  We have tried fitting our three
lightest points, our four heaviest points and our four lightest points, in
addition to fitting all five points.  In Fig.~\ref{fig:nucmphysall},
we show fits to the four heavy masses with squares, the four light 
masses with crosses
and all five masses with octagons.  The size of the plotting symbols is
proportional to the confidence level of the fit.  Fits to the three light
masses are shown with no plotting symbol since most of them have no degrees
of freedom.  (Many other
fits have such poor confidence levels that their plotting
symbols are invisible.)  When fitting the four lightest or four heaviest points
we only considered fits with up to three parameters.
We note that different fitting functions can give widely varying values.
Looking at fit 5, the linear one, we see that the errors of the fits are
small, but the result varies greatly depending upon which data is included
in the fit.  The confidence levels of these fits are unacceptable.  If one
neglects that fact,
the systematic error can
greatly exceed the quoted statistical error.  The fits with square roots
all have low values for the extrapolated mass and have large errors
because of the large derivative of the square root near zero quark mass.

We see that we have a number of octagons and two crosses of
reasonable size.  Thus, we must decide which higher order terms to use in
our fits.  In Table~\ref{tab:confidence} we give the combined confidence levels
of our fits.  Since this is a quenched simulation, all the hadron masses
on the lattices for any given coupling and volume are correlated.  We estimate
the correlation matrix from a jackknife analysis of the hadron propagators.
The simulations with different volumes or couplings are independent, so it
is easy to compute the combined confidence level of the chiral fit
by summing the $\chi^2$ and
degrees of freedom from each case and then computing the confidence level
from the sums.
For each fitting form we report two combined confidence
levels.  The column labeled ``Biggest volume'' includes the biggest volumes
at $6/g^2=5.7$, 5.85 and 6.15.  The other column also includes $N_s=20$ at
5.7 and 5.85.  We did not include the 5.54 results in this analysis because
it is quite a strong coupling, and we want our chiral fits 
to reflect the proper continuum form.

\begin{table}[tb]
\caption{Combined confidence levels of fit}
\label{tab:confidence}
\setlength{\tabcolsep}{1.2pc}
\begin{tabular}{lll}
\hline
\mystrut Fit&All 5 cases&Biggest volume\\
\hline
\multicolumn{3}{c}{\mystrut Nucleon Jackknife fits}\\
\hline
\mystrut 1 & 5.81e-273      & 1.75e-265\\
2 & 4.66e-08       & 3.42e-07\\
3 & 1.24e-01       & 8.69e-02\\
4 & 1.30e-01       & 1.05e-01\\
5 & 0.00e+00       & 0.00e+00\\
6 & 3.02e-10       & 3.25e-08\\
7 & 1.06e-24       & 6.88e-17\\
8 & 1.83e-01       & 1.99e-01\\
9 & 1.05e-02       & 2.28e-03\\
10 & 9.95e-02       & 6.25e-02\\
11 & 1.52e-07       & 1.85e-07\\
12 & 8.38e-02       & 1.51e-01\\
\hline
\multicolumn{3}{c}{\mystrut Rho Jackknife fits}\\
\hline
\mystrut 1 &0.00e+00& 0.00e+00\\
2 &5.81e-33& 2.60e-33\\
3 &1.66e-02& 7.46e-03\\
4 &1.91e-02& 6.92e-03\\
5 &0.00e+00& 1.26e-282\\
6 &1.19e-05& 1.65e-06\\
7 &2.08e-04& 1.66e-04\\
8 &4.82e-02& 2.57e-02\\
9 &1.73e-12& 8.15e-14\\
10 &4.58e-03& 2.57e-03\\
11 &1.05e-07& 1.10e-08\\
12 &5.89e-02& 3.54e-02\\
\hline
\end{tabular}
\end{table}

What we find from our table of confidence levels, is that for the nucleon,
five fits have reasonable confidence levels.  
These are the forms that contain four free parameters.
The best fit, number 8,
includes both $m^{3/2}$ and $m^2$, two terms that come into ordinary chiral
perturbation theory.  However, with a confidence level of 0.18, it is not
markedly better than fits 3 or 4 that have confidence level 0.12 and
0.13, respectively.  Both these fits have an $m^{1/2}$ term that is
characteristic of $Q\chi PT$.  Fits 10 and 12
round out the set of five fits.  Fit 10 has two terms that only appear in
quenched chiral perturbation theory.  Fit 12 has terms from ordinary
chiral perturbation theory.

The fits for the rho are more problematic.  None of the fits does an 
acceptable job fitting all the results.  In neither column is any combined
confidence level above 10\%.  
As for the nucleon, the four parameter fits, numbers
3, 4, 8, 10 and 12 seem to do the best job.
(The logarithm in fit 10 is not supposed to occur even in $Q\chi PT$.)
We have adjusted some of our rho propagator
fitting ranges to be further from the source.  This should help to avoid
bias from not being in the asymptotic region, and it also increases the
error in the particle masses, which should help in our chiral fits.
Nonetheless, we do not get good chiral fits.  Whether this is a statistical
fluctuation, an indication that we are underestimating our errors,
a manifestation of the fact that the rho is an unstable particle, or some other
problem, we have not yet determined.

It is unfortunate that we cannot determine from the confidence levels of
the fits, which is best.  It is particularly important to determine whether
or not $m^{1/2}$ should be included  
since it makes a big difference in the extrapolated value.  Of
course, it is also an artifact of quenching.
\begin{figure}[thb]
\epsfxsize=0.99 \hsize 
\epsffile{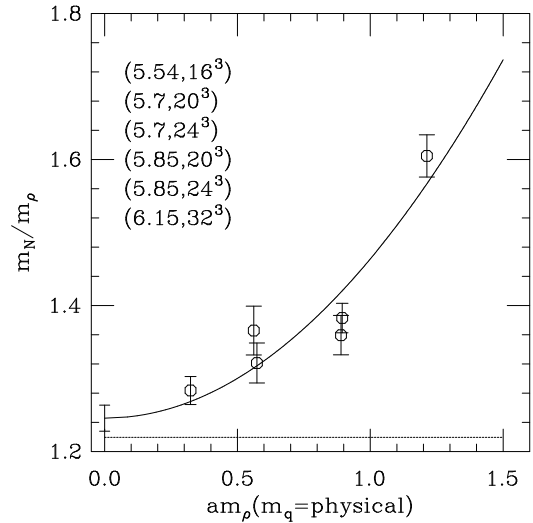} 
\vspace{-28pt}  
\caption{$m_N/m_\rho$ at the physical quark mass {\it vs}.\ $am_\rho$.  
The horizontal line is drawn at the physical value.  The error bar at
$am_\rho=0$ shows the error in the extrapolated value.}
\label{fig:ratio0808vsa}
\end{figure}

\begin{figure}[thb]
\epsfxsize=0.99 \hsize 
\epsffile{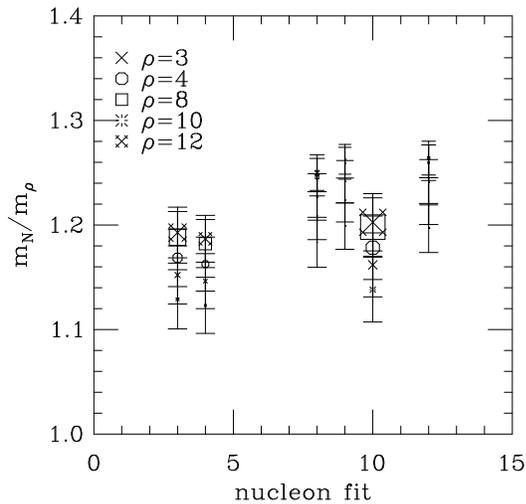} 
\vspace{-28pt}  
\caption{$m_N/m_\rho$ extrapolated to $a=0$ {\it vs}. the fit used in the
chiral extrapolation.  The size of the plotting symbol is proportional to
the confidence level of the fit to the $a$ dependence.}
\label{fig:ratiovsfit2}
\end{figure}

\subsection{Lattice spacing dependence}
The next issue to deal with is the extrapolation to
zero lattice spacing.  Because of the difficulty in deciding which chiral
fit is best, we have considered all five of the favored
possibilities for both rho
and nucleon.  We should mention that although the pion fits don't always have
acceptable confidence levels, the errors in the pion mass are so small 
($\approx <10^{-4}$) that even a fit with poor confidence level 
can be very close to our values.

For Kogut-Susskind quarks with the Wilson gauge action, both parts of the 
action have error of order $a^2$.  We fit the $a$ dependence to the
form
$$
C + b a^2 \,.
$$
In Fig.~\ref{fig:ratio0808vsa}, we show one example of the 
extrapolation in $a$ where we
have picked fit 8 for both rho and nucleon.  The octagons are for the
extrapolation to the physical quark mass as determined by $m_\pi/m_\rho$.
Only the largest volumes are used to fit the $a$ dependence.  
The confidence level of the fit is 5.5\%.
In Fig.~\ref{fig:ratiovsfit2}, we show the 
extrapolated value of $m_N/m_\rho$.  We have used different 
plotting symbols for the rho fit and plotted the ratio as a function of the
nucleon fit.  We see that nucleon fits 3, 4 and 10 which have an $m^{1/2}$ term
are systematically lower than the other fits.  There is also a systematic
pattern to the rho fits.  From largest to smallest ratio, the order is always
fit 12, fit 8 (which are nearly indistinguishable), fit 4, fit 3 and fit
10 (which we could refuse to consider on theoretical grounds).  
At the conference, two versions of this plot were shown.  The first had
equal size symbols for each fit, and makes it easy to see the systematics
described above.  The second version, shown here, has symbol sizes that are
proportional to the confidence level of the lattice spacing extrapolation.
To have a believable calculation, the chiral and lattice spacing
extrapolations must both have good confidence levels.  We see that fits 8,
12 and 4 are favored for the rho extrapolation and that the nucleon fits
that include the term $m^{1/2}$ are favored.  The fits fall close to the
experimental value of 1.22, with the ($a$ extrapolation)
favored fits falling just slightly below the experimental value.  
For any particular
set of fitting functions, the error in $m_N/m_\rho$ is about 0.02;
however, the variation of all the fitting functions is about five
times as large.  The fits that include a $Q\chi PT$ $m^{1/2}$ term  all fall
below the experimental value.

\begin{table}[tb]
\setlength{\tabcolsep}{1.5pc}
\caption{Lattices used by the IBM group}
\label{tab:gf11lats}
\begin{tabular}{lccc}
\hline
\mystrut $6/g^2$&size&\#\\
\hline
\mystrut $ 5.7 $&   $8^3 \times 32$&          2439 \cr
 $ 5.7 $&   $16^3 \times 32$&          219 \cr
 $ 5.7 $&   $24^3 \times 32$&          92 \cr
\noalign{\smallskip}
 $ 5.93 $&   $24^3 \times 36$&          210 \cr
\noalign{\smallskip}
 $ 6.17 $&   $32^2 \times 30\times 40$&          219 \cr
\hline
\end{tabular}
\end{table}

\section{WILSON SPECTRUM}
\subsection{Motivation}
The most extensive calculation (at least before this conference)
of the quenched Wilson spectrum was carried out recently on IBM's GF11
parallel computer by Butler {\it et al}.\ \cite{BUTLER}.
They attempted to control all the sources of systematic error we discussed
above.
Their ensemble of lattices is displayed in Table \ref{tab:gf11lats}.
They had three lattice sizes at $6/g^2=5.7$, but only one at their other
two couplings.  They also claimed on the basis of their calculations at
5.7 that they could make a finite volume correction to their final answer.
They obtained $m_N/m_\rho= 1.278 \pm 0.068$ at finite volume, but with the
volume correction, they got $1.216\pm0.104$, in excellent agreement with
the observed value 1.222.  This celebrated result was based on a
combination of three small sink sizes.  
The sink size parameter $r_0$ determines the size over which quark
fields are smeared.  First, the gauge is fixed to Coulomb gauge, then
the smeared field $\phi_{r_0}$ is given by
$$
\phi_{r_0}(\vec x,t) = N \sum_{\vec y} \exp( -|\vec x - \vec y|^2/r_0^2) 
  \psi(\vec y, t)
$$
where $N$ is a normalization factor.  (Similarly, a Gaussian distribution
can be used to create the source.)
They also quote results for a single sink size of 4.  
The corresponding
results $1.328 \pm 0.085$ and $1.330\pm 0.131$
show little finite volume effect and lack the impressive agreement with
experiment.  In view of the important role of the finite volume correction
and the fact that we have a larger ensemble of lattices (except for
$N_s=8$), we decided to do a Wilson spectrum calculation on our 
$6/g^2=5.7$ lattices
with $N_s=12$, 16, 20 and 24.
At the time of the conference, we had completed running on the lattices
with $N_s=16$ and 20.  We had only 150 $N_s=12$ and 45
$N_s=24$ lattices done.  All of the running has now been completed, but in
the interest of historical accuracy we will only display results that were
available during the conference (with one exception).

\subsection{Finite size effects}
The best available results for the nucleon and rho mass at $6/g^2=5.7$ were
summarized in Ref.~\cite{SGSPECREVIEW}, and the graphs also appear as
Figs.~10 and 11 at the WWW site mentioned in the introduction.  
Briefly,
the two lightest $\kappa$'s at which there are results for both $N_s=16$ and
24 correspond to $m_\pi/m_\rho=0.69$ and 0.50.  Only for the lightest quark
mass is there a finite size effect.  For the nucleon, it is a 4.8\% or
2.6 $\sigma$ effect.  For the rho, it is 3.3\% or 2.5 $\sigma$.  This is
certainly a bigger effect than is seen with KS quarks, however; the Wilson
masses are all smaller, and if the rho mass is used to set the scale, we
would say at $N_s=16$ (24) the box size is 2.3 (3.4) fm.
These differences are based on masses fit with three different sink sizes 0,
1 and 2.

\begin{figure}[thb]
\epsfxsize=0.99 \hsize
\epsffile{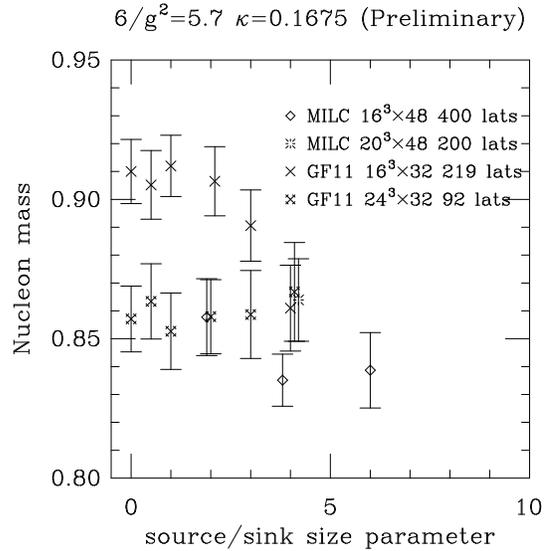} 
\vspace{-28pt}  
\caption{The nucleon mass as a function of source or sink size for $N_s=16$,
20 and 24 with $6/g^2=5.7$.}
\label{fig:kappa1675}
\end{figure}

\begin{figure}[thb]
\epsfxsize=0.99 \hsize
\epsffile{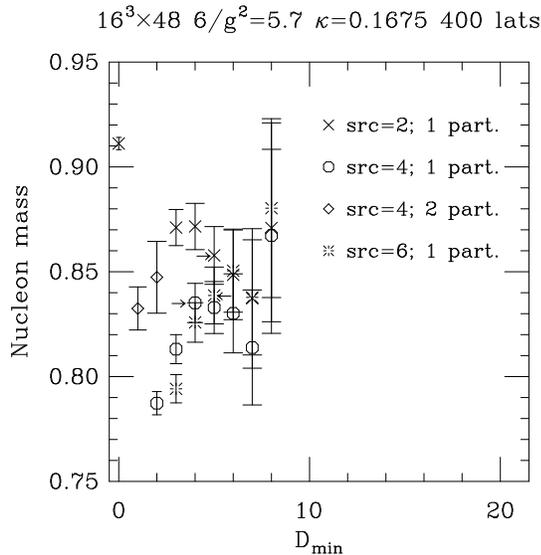} 
\vspace{-28pt}  
\caption{The nucleon mass as a function of minimum distance from source.
One and two particle fits are shown.  Arrows point to the best fits.}
\label{fig:nuc16kappa1675}
\end{figure}

Obviously, the lightest quark mass plays a very important role in the finite
size effect seen by the IBM group.  In Fig.~\ref{fig:kappa1675}, we compare our
results for the nucleon mass at $N_s=16$ with the IBM group's results at
$N_s=16$ and 24.  The mass is shown as a function of the source or sink
size parameter.  The values from fitting sizes 0, 1 and 2 simultaneously
are plotted at size 1/2.  We also include a value for $N_s=20$ that was not
available during the conference.
These results suggest that the IBM
group was really seeing a source size dependence, not a finite size effect.
In fact, a source size dependence is not a physical effect.  It can result
from not being at a large enough distance from the source to be in the
asymptotic regime.  Thus, the details of which fits were picked can be
important.  In Fig.~\ref{fig:nuc16kappa1675}, we detail
our nucleon mass fits.  We fit the nucleon propagator from $D_{min}$
to 10 for $\kappa=0.1675$.  Except for source size 4, we show only single
particle fits.  For the larger and smaller sources, we attempted some
two particle fits, but many did not converge.  (As we continue our analysis,
we will attempt these again.)  Arrows point to the fits that we selected.
In each case, we picked the best fit with more than one degree of freedom.
The confidence levels are 0.285, 0.421 and 0.427 for source size 2, 4 and
6, respectively.  Source size 2 approaches its asymptotic value from
above.  The fit with $D_{min}=3$ has confidence level 0.282, so we narrowly
missed picking it and having a much larger difference in the masses from the
three sources.  On the $16^3\times 32$ lattice, the
IBM group used the range 5--8 for sink sizes 0, 1 and
combination 0, 1 and 2.  For sink size 2, they used 3--7 and for sink sizes
3 and 4, they used 2--5.  On the larger lattice, they were a little further
from the source, using $D_{min}=4$--6.  Even with these details it is
not clear why our results and those of the IBM group differ as much as
they do.

\begin{figure}[thb] 
\epsfxsize=0.99 \hsize 
\epsffile{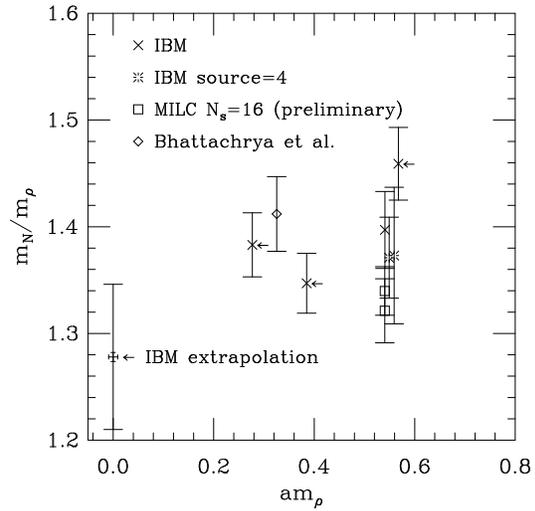}  
\vspace{-28pt}     
\caption{
$m_N/m_\rho$ {\it vs}.\ $am_\rho$ for Wilson quarks.
}
\label{fig:ratiovsmrhotsukuba} 
\end{figure}

\subsection{Chiral extrapolations}
The reader has just seen that the chiral extrapolation was a major concern
for our KS quark calculations.  For the IBM group's calculation, chiral
extrapolations were based on linear fits to the three lightest quark
masses.  That corresponds to the range $0.5 \le m_\pi/m_\rho \le 0.69$.
In our calculations, we have used the same six quark masses on all volumes.
For $N_s=16$ and 24, we have hadrons with non-degenerate quarks.  For the
mesons, we have 21 combinations of masses.

At the time of the conference, we had just done some simple extrapolations
based on the degenerate mass states.  Recently, several groups 
have studied the nonlinear terms that might come into the chiral 
extrapolations.  At Lattice '95, Sloan emphasized that the SCRI
calculations with an improved action required nonlinearities for the chiral
fits \cite{SLOAN}.  
This was also discussed by Bhattacharya {\it et al} \cite{LANL}.  
But in the latter work, they were not able to
distinguish between different higher order terms that appear in $\chi PT$,
and they did not use the full covariance matrix of the hadron masses in
doing the chiral fits, so no reliable confidence level was available.
Our fits will be done using the full covariance matrix, just as we did for
the KS spectrum.  Of particular note is another contribution to this
conference by the CP-PACS collaboration, 
presented by T.~Yoshi\'e \cite{CPPACS}.
In this very high statistics calculation with five quark masses, the
nucleon is fit with terms up to cubic in the quark mass and the rho is fit
with a quadratic.

\subsection{Lattice spacing dependence}
With only three values of the gauge coupling, the IBM group's results for
the nucleon to rho mass ratio are quite sensitive to each point.  If any
single point were removed, the result extrapolated to $a=0$ would be quite
different.  In Fig.~\ref{fig:ratiovsmrhotsukuba}, we display the results of the
IBM group, and the point from Ref.~\cite{LANL} for $6/g^2=6.0$.  
We also show our preliminary values at 5.7 for $N_s=16$.
There are many points corresponding to the $6/g^2=5.7$ and they have been
spread out slightly in $am_\rho$ to make it easier to distinguish the points.
For the IBM group, we show not only the results for sinks 0, 1 and 2, but
their reported result for sink size 4.  The two crosses to the right are
their multiple sink mass at $N_s=16$ (upper) and $N_s=24$ (lower).  The two
bursts are their sink 4 results for the two volumes.  The squares are our
preliminary results for source size 4 showing two ways of doing the chiral
extrapolation for the nucleon.
The diamond is from Ref.~\cite{LANL} where there was also some difficulty in
getting the same mass from different sources.  The masses used for this point
come from a weighted average of three different types of source or sink 
combinations.
The fancy plus is the IBM
extrapolated value based on the three crosses with arrows pointing
at them.  Bhattacharya \etal\ prefer an extrapolated value close to 1.4.

We are reluctant to redo the $a$ extrapolation at
this point.  We would like to finish the analysis of our other volumes
and do a more careful job of the chiral extrapolation.  Further, the
talk by Yoshi\'e \cite{CPPACS}
that followed this one presented evidence that the
nucleon mass
is ``much smaller'' than in previous calculations.  The couplings used
are 5.9, 6.10, 6.25 and 6.47.  As these results are likely to supersede
prior results, and our small nucleon to rho mass ratio may smoothly
join on to their weaker coupling results, it seems unwise to try to convince 
the reader that the continuum limit
quenched Wilson nucleon to rho mass ratio is nearly 1.4, especially when 
we have evidence that the quenched KS ratio is very near
or below the physical value 1.22.  (The quenched results for the
two types of lattice quarks may not agree with
experiment, but we certainly expect them to agree with each other 
in the continuum limit.)
In fact, though  CP-PACS did not present a graph with the same axes as
Fig.~\ref{fig:ratiovsmrhotsukuba}, their
results for the nucleon mass in the continuum limit are falling below
the physical value.  From their graph of the spectrum, we estimate that 
they find $m_N/m_\rho=1.17\pm 0.05$.  This would be in 
very reasonable agreement with our KS quark results.  
(See Fig.~\ref{fig:ratiovsfit2}.)

\section{CONCLUSIONS }
We have summarized some of our recent results for the quenched spectrum with
Kogut-Susskind and Wilson quarks.  For the Kogut-Susskind case, we have studied
four couplings, with five masses at each coupling.  We have found that the
chiral extrapolation is the most difficult issue to deal with.  Just on 
the basis of goodness of fit, we have not been able to confirm or refute the
presence of artifacts predicted by quenched chiral perturbation theory.
If we had been able to determine which chiral extrapolation is correct,
our error would have been about 0.02 for the nucleon to rho mass
ratio.  Without that determination, the systematic error is about five times
as large and the ratio is consistent with the experimental value.

We have also done a modest calculation with Wilson quarks at a single
coupling, in which we have
concentrated on understanding the finite size effects and the systematics
of the calculation of Ref.~\cite{BUTLER}.  Our preliminary results show
no evidence for finite size effects for $N_s=16$ with $6/g^2=5.7$.  
Our different source sizes on the $N_s=16$ lattice give consistent
nucleon masses.  After the conference, we completed our running on
$N_s=12$, 20 and 24.  We have a great deal of analysis to do to extract the
hadron masses.  Since we have six different quark masses and propagators
with non-degenerate quarks, we hope that we will be able to better elucidate
the chiral limit than we were able to do for the KS quarks.

\section*{ACKNOWLEDGEMENTS}
This work was supported by the U.S. Department of Energy under contracts
DE-AC02-76CH-0016,
DE-AC02-86ER-40253,
DE-FG03-95ER-40906,
DE-FG05-85ER250000,
DE-FG05-96ER40979,
DE-2FG02-91ER-40628,
DE-FG02-91ER-40661,
and National Science Foundation grants
NSF-PHY93-09458,
NSF-PHY96-01227,
NSF-PHY91-16964.
Computations were performed at the Oak Ridge National Laboratory
Center for Computational Sciences, the Pittsburgh
Supercomputing Center,  the National Center for Supercomputing
Applications, the Cornell Theory Center, the San Diego Supercomputer
Center, and Indiana University.
Two of us, (S.G. and D.T.), are very grateful to the Center for
Computational Physics at the University of Tsukuba for its generous
support and warm hospitality.  In addition, we thank Profs.~Y.~Iwasaki and
A.~Ukawa for all their efforts in organizing a very stimulating workshop.

\end{document}